# Site-selective functionalization of plasmonic nanopores for enhanced fluorescence and Förster Resonance Energy Transfer

*Xavier Zambrana-Puyalto, Nicolò Maccaferri, Paolo Ponzellini, Giorgia Giovannini, Francesco De Angelis, and Denis Garoli\**

Nanostructures Department, Istituto Italiano di Tecnologia (IIT), Via Morego 30, 16163 Genova, Italy

\*corresponding author e-mail: denis.garoli@iit.it

**Abstract.** In this work, we use a site-selective functionalization strategy to decorate plasmonic nanopores with one or more fluorescent dyes. Using an easy and robust fabrication method, we manage to build single plasmonic rings on top of dielectric nanotubes with different inner diameters. The modulation of the dimension of the nanopores allows us to both tailor their field confinement and their Purcell Factor in the visible spectral range. In order to investigate how the changes in geometry influence the fluorescence emission efficiency, thiol-conjugated dyes are anchored on the plasmonic ring, thus forming a functional nanopore. We study the lifetime of ATTO 520 and ATTO 590 attached in two different configurations: single dye, and FRET pair. For the single dye configuration, we observe that the lifetime of both single dyes decreases as the size of the nanopore is reduced. The smallest nanopores yield an experimental Purcell Factor of 6. For the FRET pair configuration, we measure two regimes. For large nanopore sizes, the FRET efficiency remains constant. Whereas for smaller sizes, the FRET efficiency increases from 30 up to 45% with a decrease of the nanopore size. These findings, which have been also supported by numerical simulations, may open new pathways to engineer the energy transfer in plasmonic nanopores with potential applications in photonics and biosensing, in particular in single-molecule detection towards sequencing.

In recent years, plasmonic nanopores have been proposed for applications in single molecule detection towards sequencing[1]. Compared to the more common solid-state nanopores which are typically used for single molecule experiments based on ionic current measurements[2], plasmonic nanopores offer interesting advantages for optical spectroscopic approaches[3,4,5]. Some of the key features of plasmonic nanopore are reduction of detection volume, localization of the electromagnetic field and increase of the signal-to-noise ratio. As a result, plasmonic nanopores find application in single molecule detection and sequencing based on Surface Enhanced Raman Scattering (SERS)[4], as well as in fluorescence spectroscopy experiments[6,7]. In particular, enhanced fluorescence in plasmonic nanopores has been verified in single molecule DNA detection[8,3]. In these works, one or more nucleotides are tagged with a suitable dye, and their enhanced emission during the nanopore translocation process enables the fingerprint recording of the molecule. It is worth mentioning that fingerprinting of a single molecule has been recently proposed for another important application, i.e. protein sequencing[9]. In a recent article, van Ginkel and co-workers sequentially read out the signals from a single protein by means of the the Föster Resonant Energy Transfer (FRET) between a dye fixed on a nanopore and the dyes attached to the protein[10]. Despite not using a plasmonic nanopore, this proof-of-concept experiment shows the potentiality of using FRET to carry out single molecule sequencing using a nanopore. As far as FRET experiments using plasmonic nanostructures are concerned, in 2014 Ghenuche *et al.* demonstrated that a plasmonic nanohole can produce a six-fold increase in the single molecule FRET rate[11]. Furthemore, in 2016 de Torres *et al.* showed that plasmonic antennas can enable forbidden dipole-dipole Föster energy transfer exchanges[12]. Despite the significant growing interest on this topic, there have not been many works on FRET experiments using plasmonic nanopores. In addition, no flow-through experiments making use of FRET effects have been reported so far. One of the key aspects to perform FRET-based sequencing experiments is the immobilization of one of the FRET pair dyes close to a plasmonic surface. In this respect, surface functionalization protocols for dye grafting have been extensively investigated[13], but the selective functionalization at the specific positions where the plasmonic nanostructure is present is still challenging and just a few works have been reported on that[14–16]. Notice that a proper site-selective functionalization significantly reduces the amount of the background signal coming from the substrate as well as it facilitates the investigation of fluorescent emission from specific isolated points. Clearly, fundamental studies to optimize the use of plasmonic nanopores are still missing.

In this work, we show that it is possible to achieve an easy site-selective decoration of metallic nanopores in large arrays with thiol-modified dyes. Our functionalization method has a great control over the nanoplasmonic sites to be functionalized, and it is done in one single step. We use this functionalization method to experimentally characterize the emission efficiency of plasmonic nanopores of four different sizes. The characterization is carried out for single fluorophore and donor/acceptor (D/A) FRET configurations. The experimental evidence is strongly supported by numerical simulations. In particular, we show that smaller nanopores produce higher field enhancement and higher fluorescence emission rates. We observe that the D/A pairs have a FRET efficiency of 30% for large nanopores, and we measure that this efficiency increases when the size of the nanopore shrinks. We believe that these configurations can find important use in single molecule flow-through experiments for single molecule detection towards sequencing.

**Results and discussion**

We have fabricated plasmonic nanopores by building a gold ring on top of a dielectric hollow pillar. The fabrication procedure, described in the Methods section, allows us to obtain different diameters of the hole, spanning from few tens to hundreds of nm. Representative SEM images of illustrative samples are displayed in Figure 1. In our experiments, four different geometries have been investigated, each of them corresponding to a different ionic current used for the fabrication (see Methods and Figure 1). The pillars have been prepared in 30x30 micron arrays, with a distance of 5 micron between adjacent pillars. The four representative samples, defined by the dimensions of their inner ($D_{in}$) and outer ($D_{out}$) diameters of the plasmonic holes are given in Table 1.

| Notation | $D_{in}$ (nm) | $D_{out}$ (nm) |
|---|---|---|
| A | 70 ± 30 | 230 ± 30 |
| B | 170 ± 30 | 310 ± 40 |
| C | 250 ± 60 | 370 ± 10 |
| D | 330 ± 40 | 500 ± 10 |

**Table 1.** Notation and average dimensions of the four different kinds of fabricated nanopores.

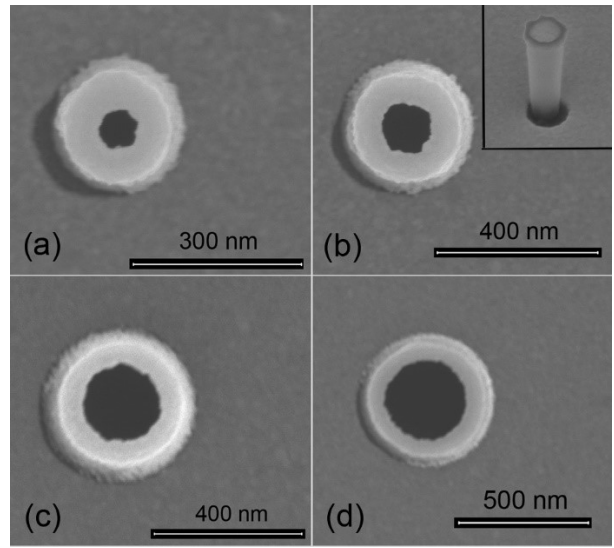

**Figure 1.** SEM images of the fabricated nanopores. (a) Nanoring prepared with 80pA ion current; (b) ring prepared with 230 pA (inset tilted view of the dielectric pillar and the metallic ring at the top); (c) ring prepared with 430 pA; (d) ring prepared with 790 pA.

We expect that different diameters correspond to different configurations of field confinement and enhancement inside the nanoholes. Finite Element Methods (FEM) simulations implemented in Comsol Multiphysics with the Radio Frequency Module have been used in order to elucidate the behaviour of the realized. In Figure 2a, we show the spectra of the four structures, calculated under a linearly tightly focused Gaussian beam. It is observed that the spectrum of the nanopores red-shifts as the inner diameter is increased. It is also seen that the average field enhancement in the ring volume is enhanced as the size is decreased. Then, in Figure 2b, we show the average field intensity enhancement $\langle |E/E_0|^2 \rangle$ calculated in the whole volume of the nanopore (see Methods). The illumination in Figure 2b is chosen to be a tightly Gaussian beam at 532 nm, same as in the experiments (see Methods for the more details). We observe that the field enhancement that the dyes feel inside the pore decreases when the size of the nanopore is increased. Also, looking at the near-field profiles in the insets of Figure 2b, it is observed that the smallest nanopore ($D_{in}$ = 70 nm) is the only one where the field is confined in the whole volume of the pore. In the other cases, the field is especially localized around the walls of the nanopore. This fact is in agreement with some previous results on similar structures[17].

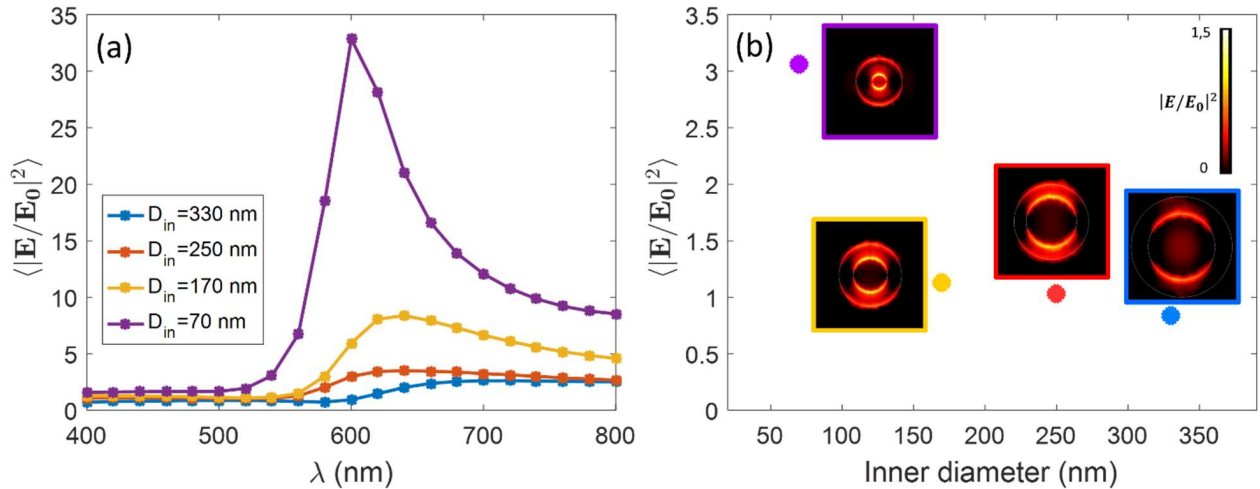

**Figure 2.** a) Spectra of the intensity enhancement for the four nanopores under a tightly focused Gaussian beam. b) Average field intensity enhancement for the four different fabricated nanopores calculated in the volume of the plasmonic ring. The simulations are done for a linearly polarized Gaussian beam that is tightly focused ($w_0$ = 350 nm) at 532 nm. Insets: near-field profiles induced by the tightly focused Gaussian beam at 532 nm.

The fabricated nanostructures comprise an isolated gold ring on top of a dielectric tube, as it can be seen looking at Figure 1 and 3(a). The choice of this configuration/geometry is indeed the key aspect for the site-selective functionalization method used here. The nanotube is used as a nanochannel where molecules can translocate/diffuse through. The proposed method, inspired by a functionalization strategy recently reported for 2D materials[18], is based on the conjugation between a gold (or another noble metal) surface and a thiolate-dye (Fig. 3). In particular, to perform a site selective decoration of metallic holes, we used a HS-PEG2000-NH$_2$ molecule as a linker between the gold surface and NHS-activated ATTO dyes. Notice that the PEG2000 molecule has a size of approximately 6 nm and lower linker lengths can be achieved using shorter PEG molecules[19]. While the protocol of functionalization of the dyes is reported in Methods, the procedure of nanopore decoration is the following (see Fig. 3-Top Panel):

1) A 50 μM solution of HS-PEG-NH-OC-ATTO dye is prepared in EtOH. 2) The plasmonic nanopores are prepared on a $Si_3N_4$ membrane. The metal is only deposited on one side of the substrate (see Methods). 3) The sample is cleaned in oxygen plasma for 180 seconds to facilitate the diffusion of the dyes through the nanochannels. The surface that is cleaned is the one where there is no metal. 4) We suspend the sample on top of an EtOH bath. The metallic face of the sample is in contact with the EtOH bath, whereas the $Si_3N_4$ face is in contact with the air. 5) We wet the dry side with a 3 μL droplet of HS-PEG-NH-OC-ATTO diluted in EtOH. Due to the different concentration, the dyes move towards the EtOH bath and they get attached to the metallic part when they try to reach it. 6) Before the droplet containing the solution with dyes starts to dry off, we add another droplet of 3 μL of the same solution. We repeat this process three times. 7) The sample is rinsed off in a second pure EtOH bath, and the site-selective functionalization is achieved. It is worth mentioning that we chose EtOH in order to avoid increase the wettability of the nanopores.

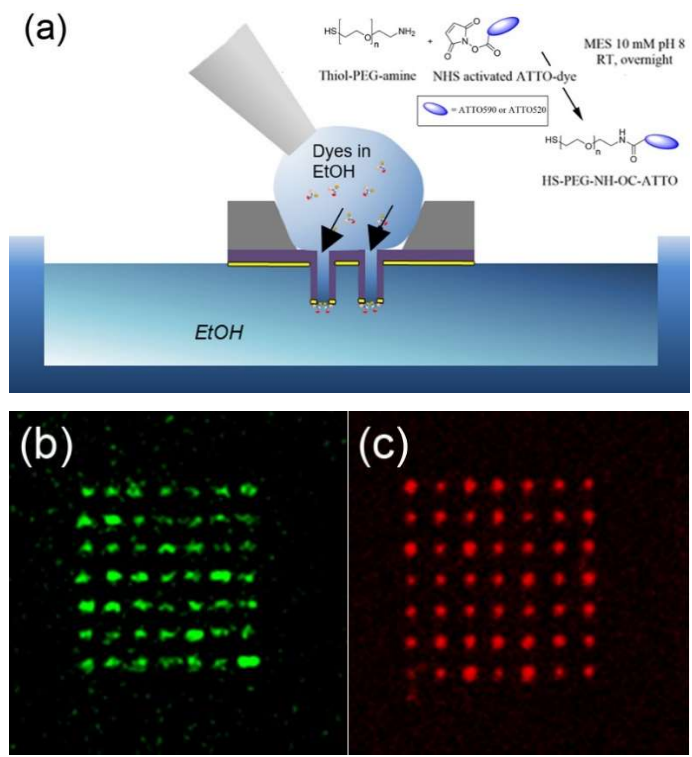

**Figure 3.** (a – Top panel)Sketch of the site-selective functionalization process. The nanopores are in contact with an EtOH bath. The other face of the sample is in contact with a solution of dyes in EtOH. The dyes diffuse through the channels and get attached to the nanopore thanks to the thiol group. Fluorescence images of plasmonic nanopores site-selectively decorated with ATTO dyes. b) The nanopores are functionalized with ATTO520. The laser excitation is at 488nm and the fluorescence detection at 500-550nm. c) The nanopores are functionalized with ATTO590. The laser excitation is at 561nm and the fluorescence detection at 570-620nm.

In order to avoid background contributions, we are interested in functionalizing plasmonic nanopores isolated from the substrate. Yet this method of site-selective functionalization can be used on any nanostructure that contains nanochannels. The thiol-terminated dyes strategy is suitable for noble metals, while other functional group can be defined in order to apply this methods also to other materials[13]. In Fig. 3(b)(c), we depict fluorescence confocal images from two different samples functionalized with ATTO520 and ATTO590 respectively. Both images have been taken with a Nikon A1 confocal microscope. In Fig. 3b(c) we show confocal fluorescence images obtained in the 500-550nm (570-620nm) range using a laser at 488nm (561nm)[1]. It can be observed that the fluorescence is localized over the array of nanopores, hence demonstrating the site-selective functionalization. We also see that the fluorescence intensity is not the same for all the nanopores. This is because our protocol does not allow us to control the number of molecules that are attached on the metallic surface. This is not an issue for this study, as the lifetime measurements that we have performed are independent of the number of molecules. Thus, we leave the control of the number of molecules attached during the functionalization process as future work.

After verifying that our site-selective functionalization method works properly, we set out to study the emission properties of ATTO520 and ATTO590 dyes. As mentioned earlier, we have measured the emission properties of the dyes in two different functionalization configurations: single dye and FRET (D/A) pair. For each of the two functionalization configurations, we have measured four different arrays of antennas. Each array is composed of 49 nanopores, each of them having a different size. The mean internal/external diameter ($D_{in}/D_{out}$) has been given in Table 1.

The lifetime measurements have been carried out using a pulsed laser at 532nm. We have used two channels of detection. One channel is set up to detect the fluorescence from the ATTO 520 dyes in the 553-577 nm spectral band. The other channel detects the fluorescence emitted in between 593-643 nm. The protocol that we have followed to measure the lifetime of the ATTO dyes as well as the details about the optical set-up are given in Methods. In Figure 4(c), we have displayed representative measurements of the lifetime traces that we have obtained for the four structures functionalized with ATTO 520 dyes. Moreover it has also been included a reference lifetime measurement of ATTO 520 dyes thiol-functionalized on top of the gold flat surface of the membrane. In Figure 4(a), we display the lifetime measurements that have been obtained after functionalizing the four different fabricated arrays of nanopores with single ATTO 520 and 590 dyes. Each data point in Figure 4(a) is the result of averaging out approximately 100 lifetime measurements (see Methods). It is observed that both dyes follow a similar lifetime vs nanopore size trend. That is, the lifetime of both dyes decreases as the inner diameter of the nanopores shrinks.

---

[1] Note that for lifetime measurements we used a different setup with a different illumination.

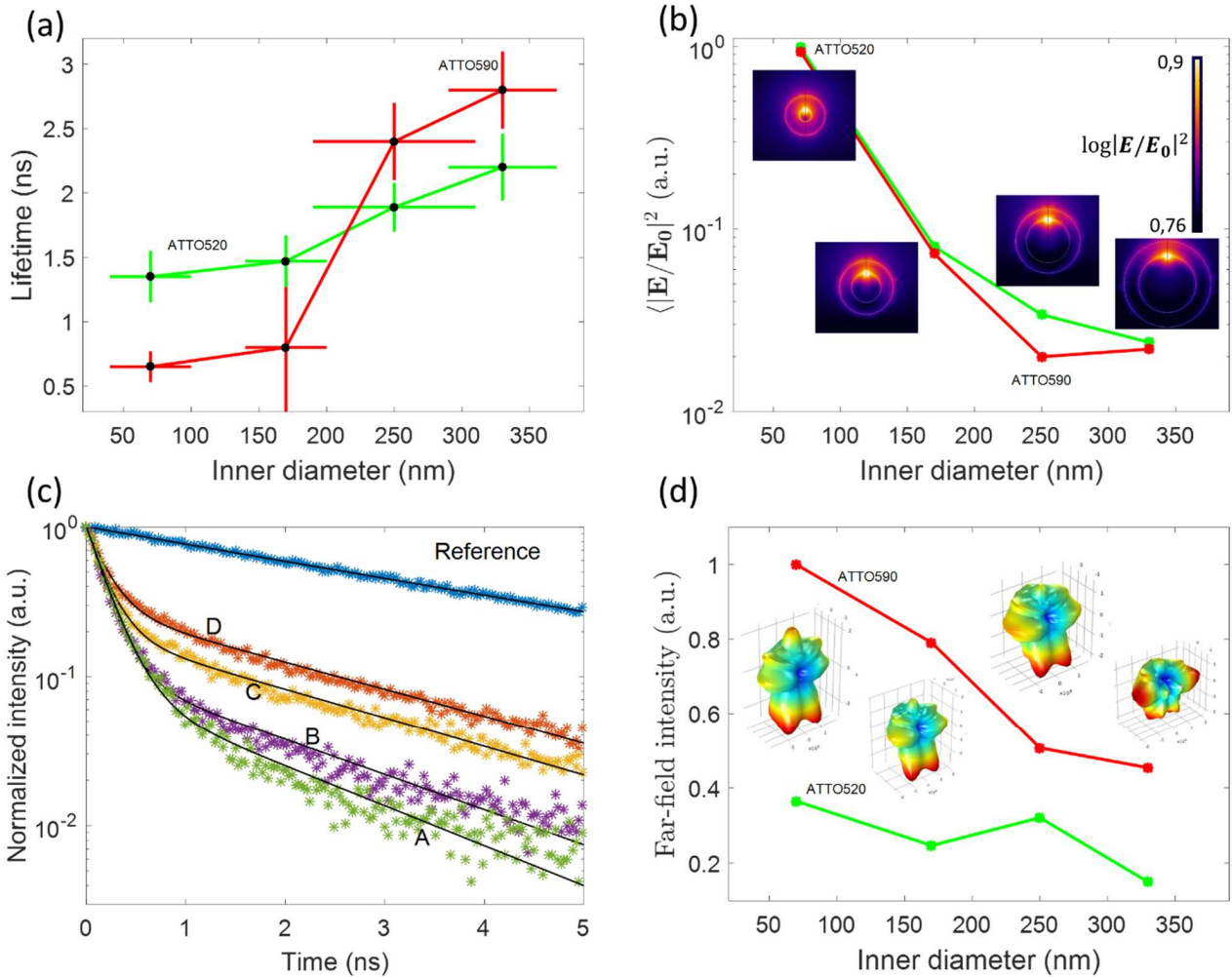

**Figure 4.** (a) Lifetime measurements of site-selective functionalized plasmonic nanopores with single ATTO 520 and ATTO 590 as a function of the inner diameter of the nanopore. Each data point is the result of the measurement of two arrays of 49 nanopores (see Methods). The fluorescence from the ATTO 520 (590) dyes has been measured in the 553-577 (593-643) nm spectral band. (b) Average intensity emitted by an electric dipole in the nanopore volume. The dipoles emit at 565nm and 618nm respectively. The insets are near-field distributions of the field induced by the dipole emitting at 565nm. (c) Representative lifetime measurements of the four nanopores, as well as the reference case. (d) Far-field intensity emitted by dipoles emitting at 565 nm and 618nm. The far-field is calculated on a horizontal plane on top of the nanopore. The insets are the far-field distributions induced by the emitting dipole at 618 nm.

This phenomenon is quite intuitive: when the size of the nanopores is smaller, it is more likely that the emission of the dyes gets reflected off the plasmonic nanostructure and absorbed back by the emitter, thus enhancing the emission rate. In contrast, understanding why the lifetime reduction is more drastic for the ATTO 590 than for the ATTO 520 is less intuitive. We have carried out some simulations to see how the emission of an electric dipole emitting at 565nm (ATTO 520) and at 618nm (ATTO 590) is like. We have placed the dipole at 6 nm of the inner wall of the nanopore (to account for the PEG molecule), and computed the intensity inside the nanopore (see Methods). Given a unitary dipolar emission, we expect the emission to increase as the size of the pore is decreased, since the emission goes as the inverse of the lifetime[20]. This behavior is indeed well captured by the simulations in Figure 4(b). Moreover, the near-field simulations also capture the big change in the slope of the trend line that the ATTO 590 dye follows in between 170 and 250 nm. The insets in Figure 4(b) panel are the near-field distributions induced by the dipole excitation for each of the four structures. For the sake of completeness, in Figure 4(d) panel, we have displayed the far-field intensity distributions as insets, and their corresponding integrals on a plane. We see that the same general behavior (emission decreases as size increases) mostly holds.

In order to get an idea of the mean Purcell Factor (or enhancement of the spontaneous decay rate, $\Gamma/\Gamma_0$) produced by the functionalized nanopore, we have measured the lifetime of the HS-PEG-NH-OC-ATTO dyes deposited on top of the gold-coated membrane. That is, the dyes are spotted and then anchored in the same membrane as the plasmonic nanopores, but out of their influence. In Table 2 we present the results of these measurements, as well as the corresponding Purcell factors for all the nanopores and the two dyes. It is observed that the lifetime of both dyes on top of a golden membrane is of the order of 3.5 ns. Given these lifetimes, the nanopores yield Purcell factors in between 1 and 6, which are in accordance with those measured for similar structures[21].

| Dye | Lifetime (ns) | $\Gamma/\Gamma_0$ for A | $\Gamma/\Gamma_0$ for B | $\Gamma/\Gamma_0$ for C | $\Gamma/\Gamma_0$ for D |
|---|---|---|---|---|---|
| ATTO 520 | 3.5 ± 0.2 | 3.2 ± 0.6 | 2.5 ± 0.4 | 1.9 ± 0.2 | 1.6 ± 0.2 |
| ATTO 590 | 3.6 ± 0.1 | 6 ± 1 | 5 ± 3 | 1.5 ± 0.2 | 1.3 ± 0.1 |

**Table 2.** Lifetime measurements of ATTO dyes dried off on a membrane.

We have also characterized the nanopores in the FRET configuration. That is, we have attached both ATTO dyes at the inner wall of the nanopores using our site-selective functionalization. The concentration used for the functionalization has been halved with respect to the concentration used for the single dyes. Notice that the FRET effects that are measured here are mean effects, as the distance between the D/A pairs cannot be controlled in an efficient manner. In Figure 5(a), we depict the lifetime of the ATTO 520 dyes both in single and FRET configuration. In Figure 5(b), we plot the FRET efficiency obtained using the data in Figure 5(a). The efficiency is obtained as $\eta_{FRET} = 1 - \tau_{DA}/\tau_D$ (see Methods). These two plots give us different information. Firstly, in Figure 5(a) we observed that the presence of the acceptor shortens the lifetime of the donor, which is one of the trademark effects of FRET interaction. Then, in Figure 5(b) we see that the FRET efficiency is almost constant for large nanopore sizes (30% and 31%), and then it increases as the size of the nanopore shrinks. Despite the fact that the distance between the dyes is not constant, the efficiency values obtained are comparable to those obtained by some other groups using similar structures[22]. The efficiency vs size behaviour is also quite intuitive: we assume that the dyes are attached onto the wall of the nanopore at a constant rate. That is, we assume each infinitesimal area of the nanopore has the same number of dyes. When the nanopore is large, we assume that the FRET efficiency stems from the interaction between the dyes in this infinitesimal area. That is, the dyes interact with their first neighbours, but they do not interact with the dyes in the other infinitesimal areas, as the distance between them is longer than the typical 1-20 nm FRET range. This situation changes as the curvature of the inner wall increases. The dyes from each infinitesimal area are effectively approached to the rest of neighbouring areas. As a result, the interaction between D/A pairs is no longer restricted to first neighbours, thus increasing the efficiency of the FRET energy exchange. A sketch of this explanation is depicted in Figure 6, where the position of the dyes on the PEG molecule is determined by an orthogonal line to the surface of the inner wall. In this toy model, it is observed that the average distance between the FRET pair is reduced from 7.4 to 6.7 nm due to the increase of the curvature of the inner wall. This change in average distance is quite consistent with the FRET radius that we have experimentally obtained, which are $7.5 \pm 0.9$ and $6.8 \pm 0.8$ nm for the largest and smallest nanopore respectively (see Methods). That is, our functionalization method allows us to attach two dyes in the inner walls of the nanopores, and then retrieve the curvature of the nanopore via lifetime measurements.

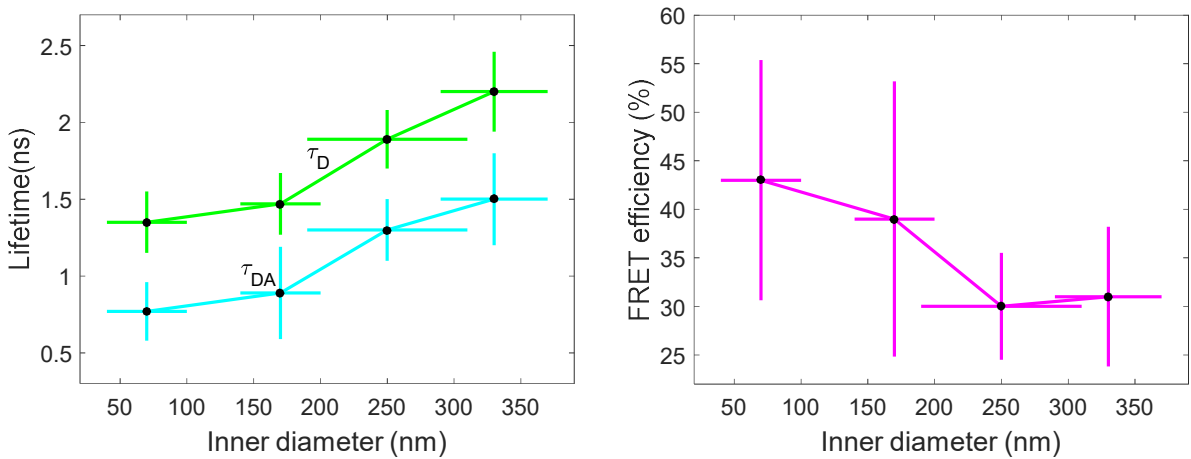

**Figure 5.** a) Lifetime measurements of site-selective functionalized plasmonic nanopores with single ATTO 520 and with ATTO 520 and ATTO 590 altogether. Each data point is the result of the measurement of two arrays of 49 nanopores (see Methods). The fluorescence from the ATTO 520 dyes has been measured in the 553-577 nm spectral band. b) Average FRET efficiency for the ATTO520-ATTO590 FRET pair. The FRET efficiency is computed with the relation $\eta_{FRET} = 1 - \tau_{DA}/\tau_D$ with the data given in Figure 6a.

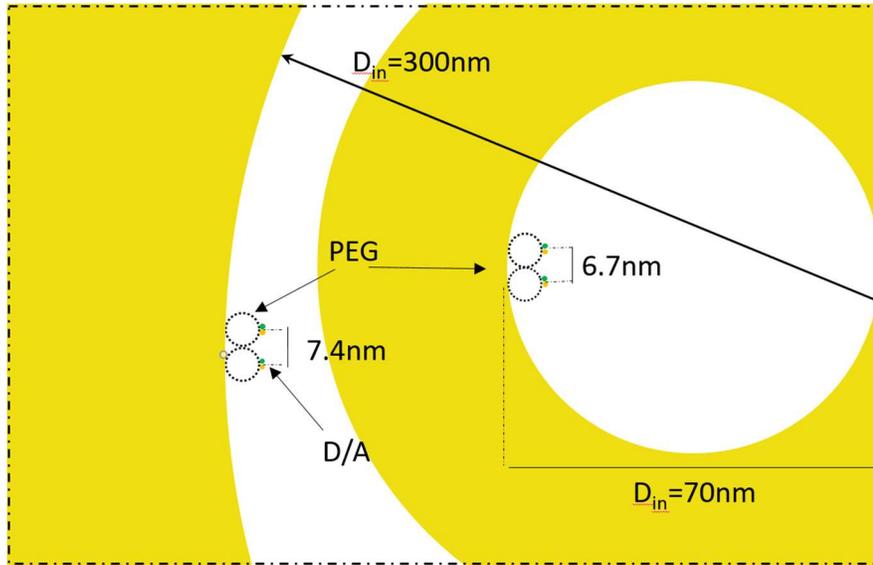

**Figure 6.** Sketch of the curvature influence on the FRET efficiency. The structure in the centre of the picture represents a nanopore with an inner diameter of $D_{in}$=70 nm. The outer yellow represents a nanopore with an inner diameter of $D_{in}$=300 nm. The PEG molecules are displayed as dashed circles of 6nm. The ATTO dyes are drawn as coloured dots at the end of the PEG molecules. An orthogonal line to the surface of the inner wall determines the position of the dyes on the PEG molecule.

**Conclusion**

In summary, we have investigated a site-selective functionalization technique to decorate metallic nanopores. The nanopores have been fabricated using a robust technique that allows us to control the dimension of inner channel of the nanopore. In order to functionalize the metallic nanopores, we have modified NHS-activated dyes with a thiolate PEG. We have attached the HS-PEG-NH-OC-ATTO dyes to the plasmonic ring using a concentration gradient in between the two sides of the sample. We have functionalized the structures for four different nanopores sizes. We have observed that the emission efficiency of the plasmonic nanostructures increases as the inner diameter of the nanopore shrinks. This trend has been tested for single dyes as well as for FRET pairs. In the case of FRET pairs, it has been observed that the efficiency of the energy exchange increases for small sizes, whereas it stays constant for large sizes. The novel approach herein presented in which plasmonic nanopores are used for fluorescence energy transfer could lead to alternative detection schemes in photonics, biosensing and sequencing.

**Methods**

*Nanopores array fabrication*

The fabrication of the metallic nanoholes follows simple and robust procedures. The substrate is a $Si_3N_4$ membrane (100 nm thick) prepared on a Silicon chip. The fabrication of plasmonic ring nanopores array follows the procedure illustrated in Ref.[23] and used by our group in several recent papers. A thin layer of S1813 optical resist is spinned on top of the membrane with a final thickness equal to the height of the structure we want to obtain. This layer is exposed by secondary electron during the FIB milling of the nanoholes from the bottom of the membrane. For this experiment, four ionic currents have been used: 80, 230, 430 and 790 pA. Smaller hole radii can be achieved by using lower ionic current still preserving the robustness of the procedure. After the exposure, the samples have been developed in acetone and rinsed in EtOH. Finally, e-beam depositions (ca. 3nm Ti and 30 nm Au) have been performed in order to deposit the ring on top of the obtained dielectric pillar. The directionality of the physical evaporation process is fundamental to avoid the deposition along the pillar wall. A deposition rate below 0.3 A/sec ensures low roughness. All the samples have been treated in $O_2$ plasma before the dyes functionalization.

*Optical set-up*

The experiments are performed on an inverted microscope with a long distance water-immersion microscope objective with NA=1.1. The light beam, which enters the microscope through its rear port, is obtained with a fibred supercontinuum laser with a central wavelength of λ=532 nm. The spectral bandwith of the laser is ±5 nm, the repetition rate is 78 MHz and the pulse width is approximately 65 ps. The power of the laser before entering the rear port of the microscope is 0.35 µW. The samples that we use for the experiment are held on a sample holder, which is attached to a micro and a nanopositioner. The nanopositioner is used to center the nanopore with respect to the incident beam. We consider that the nanopore is centered with respect to the beam when the fluorescence intensity is maximum. The nanopositioner is also used to place the sample at the z plane where the fluorescence counts are maximized. The detection of the fluorescence is done in reflection. The fluorescence is separated from the backscattering of the sample by means of three filters. Firstly, a notch dichroic filter at 532 nm and a longpass filter at 550 nm remove most of the backscattering. Then, a dichroic filter at 594 nm splits the fluorescence in two paths. The fluorescence on the path for wavelengths below 594 nm mostly belongs to the ATTO520, while the fluorescence for wavelengths longer than 594 nm is mostly due to the emission of ATTO590. A different third filter is added on each path. A bandpass filter at 565/24 nm (618/50 nm) is added on the path of detection of ATTO520 (ATTO590). An avalanche photodiode (APDs) is set up on each path to detect the fluorescence. The signal of the APDs is recorded with a time-correlated single photon counting module.

*Lifetime measurements*

The APDs as well as the laser are connected to a time-correlated single-photon counting module (Swabian Instruments) in time-resolved mode. Making use of a home-built code, we build a histogram of 300 bins, each of them having a temporal width of 30 ps. A time delay of 6.56 ns is applied to the laser channel, so that the histogram is monotonously decreasing. The histogram measurement is carried out for 25 s, and it is repeated four times, yielding four histograms. A biexponential function of the kind $Ae^{-t/\tau_a} + Be^{-t/\tau_b}$ is used to fit the decay curve for each histogram. Typically, the $Ae^{-t/\tau_a}$ exponential contains information about the backscattering (or impulse response function), whereas $Be^{-t/\tau_b}$ contains information about the fluorescence. This is easily observed, as the fluorescence signal tends to decrease over time due to the photobleaching of dyes which are illuminated for long periods close to a metallic surface. This photobleaching effect is usually captured in the fit by a decrease of the coefficient $B$. Then, measuring four histograms allows us to see if the lifetime measurement is stable over time, or not. We obtain the lifetime of a particular nanopore by doing a weighted arithmetic mean of the four lifetimes, the weights being $B/A$. The same measurement is carried out for each of the nanopores of the array (49 in total), and it is repeated for a second array. Then, the lifetime of the array is given as the average of all the significant lifetimes of the two arrays. The lifetime measurements that are not considered significant are due to the fact that their value are more than 2σ away from the average. The error of the measurement is obtained as the standard deviation associated to the average with the significant points.

*FRET measurements*

The FRET efficiency of the antennas is computed as $\eta_{FRET} = 1 - \tau_{DA}/\tau_D$, with $\tau_D$ being the lifetime of the donor by itself and $\tau_{DA}$ the lifetime of the donor when it is in the neighbourhood of an acceptor. Due to the fact that we cannot control the distances in between the donor and the acceptor, our experiments yield mean measurements of the FRET efficiency. The concentration of the donor has been maintained for both experiments, *i.e.* single donor and FRET pair. A control check with acceptor sensitization has not been performed because of our lack of control of the number of molecules and the decreasing fluorescence signal due to photobleaching. The error of $\eta_{FRET}$ is computed doing a standard analysis of the statistical propagation of the uncertainty: $s_f = \sqrt{\sum_i \left(\frac{\partial f}{\partial x_i}\right)^2 s_{x_i}^2}$.

The D/A mean distance is obtained using the expression $R = R_0 \left(\frac{\tau_{DA}}{\tau_D - \tau_{DA}}\right)^{1/6}$, where $R_0 = 6.5$ nm is Förster distance of the two ATTO dyes used in the experiment[24].

*Chemical modification of ATTO dyes*

HS-PEG2000-NH2 [mercapto poly(ethylene glycol) amine] (2,27 mg, 1,135 µmol) were solubilised in 1 mL of MES 10 mM pH 8 (1,14 mM). From this stock solution, aliquots of 100 µM of HS-PEG2000-NH2 were prepared using the same buffer, MES 10 mM pH 8.
100 µL of the HS-PEG2000-NH2 aliquot were added to a solution of ATTO590-NHS ester (100 µM) in MES 10 mM pH 8. The mixture was stirred at room temperature, at dark, overnight. The same procedure was followed to label HS-PEG2000-NH2 with ATTO520-NHS ester. Even in such case, the two reagents were used in the stoichiometric molar ratio of 10:1, respectively for the dye and PEG. The excess of dye was then removed by 24 hours of dialysis (cut-off membrane: 1 kDa). Once purified, the labelled PEG was lyophilised overnight (Lio5P, Kambic). The amount of dye effectively linked to PEG was quantified by measuring the absorbance in water of the labelled PEG and by means of the specific molar attenuation coefficient and Beer-Lambert law. The so prepared HS-PEG-NH-OC-ATTO590 and HS-PEG-NH-OC-ATTO520 were stored at -20°C in aqueous solution and diluted at the desired concentration in EtOH prior surface functionalisation experiments.

Materials: ATTO520 NHS ester, BioReagent, suitable for fluorescence, ≥80.0% (coupling to amines), ATTO590 NHS ester, BioReagent, suitable for fluorescence, ≥60% (coupling to amines), HS-PEG2000-NH2 HCl Salt (average Mn 2,000) and 4-Morpholineethanesulfonic acid (MES) hydrate, ≥99.5%, were purchased from Sigma Aldrich. MES 10 mM pH 8 was prepared as followed: a stock solution of 100 mM MES was prepared solubilising MES hydrate (9,76 g, 50 mmol) in of DI water (500 mL); 10 mL of this solution were diluted with 90 mL of DI water reaching 10 mM as final concentration; the pH was adjusted to pH 8 adding 0,1 M NaOH dropwise. SpectrumTM Spectra/Por® 6 Standard RC Pre-wetted Dialysis Tubing MWCP 1 kD were purchased from Fisher.

*Electromagnetic Simulations*

Numerical simulations based on the Finite-Element Method implemented in the RF Module of Comsol Multiphysics® were carried out to investigate the electromagnetic response of an isolated nanopore. The dimensions of the simulated structures were set according to the average sizes obtained from SEM measurements. A dielectric constant n = 1.33 was used for water, and n = 1.5 was set for the glass substrate. The refractive index of Au was taken from Rakic[25]. The model computes the electromagnetic field in each point of the simulation region, enabling the extraction of the quantities plotted along the manuscript. The unit cell was set to be 1500 nm wide in both x- and y-directions and 2000 nm along the z-direction, with perfect matching layers (200 nm thick) at the borders. A linearly polarized Gaussian beam is tightly focused down to a spot size of $w_0 = 350$nm. The beam impinges on the top side of the structure, namely where the nanopore is placed, from the water side. The field intensity enhancement is computed as the average of field intensity in the volume of the nanopore. We define the volume as $\pi D_{in}^2 t/4$, with *t = 30* nm being the width of the deposited gold ring.

For the simulations with the dipoles, a dipole oscillating along the in-plane direction of the nanopore at either 565 nm or 618 nm was placed at 6 nm from the inner wall of the nanopore for all four the cases treated in the manuscript. In this case the dipole is the only field source within the simulation region. The field intensity enhancement is also computed across the whole volume of the pore.

The far-field, *i.e.* the scattering amplitude of the structure in the momentum space, is calculated from the near field using the Stratton-Chu formula[26], namely $E_{far} = \frac{ik}{4\pi} r_0 \times \int [\boldsymbol{n} \times \boldsymbol{E} - \eta r_0 \times (\boldsymbol{n} \times \boldsymbol{H})] e^{ikr \cdot r_0} dS$ , where $\boldsymbol{E}$ and $\boldsymbol{H}$ are the fields on the surface S enclosing the structure, $\boldsymbol{r}$ is the radius vector (not a unit vector) of the surface $S$, $E_{far}$ is the field calculated at the point $\boldsymbol{r_0}$, $\boldsymbol{n}$ is the unit normal to the surface $S$, $k$ is the wave number and $\eta = \sqrt{\mu/\varepsilon}$, is the impedance. Then, the far-field calculations carried out in Figure 4d are computed as an average of the far-field intensity projected on a plane at infinity. The plane is located in the semispace where the water-immersion objective is located.

**Acknowledgements**

The research leading to these results has received funding from the European Research Council under the Horizon 2020 Program, FET-Open: PROSEQO, Grant Agreement n. [687089]. X.Z.-P. wants to thank Giorgio Tortarolo and Daniel Price for interesting discussions, as well as Marco Scotto and Michele Oneto for their help with the confocal microscope.

**Author Contributions**

XZP performed the optical measurements and the site-selective functionalization, DG and PP prepared the samples, NM performed numerical simulations, GG performed the dyes functionalization, FDA and DG coordinated the work. All the authors contributed to the manuscript preparation and revision.